\documentstyle[elsart12,epsfig]{elsart}

\begin{document}

\begin{frontmatter}

\title{On the reliability of negative heat capacity measurements}

\author[bologna]{M.~D'Agostino},
\author[lpc]{R.~Bougault},
\author[lpc]{F.~Gulminelli},
\author[bologna]{M.~Bruno},
\author[bologna]{F.~Cannata},
\author[GANIL]{Ph.~Chomaz},
\author[lnl]{F.~Gramegna},
\author[mi]{I.~Iori},
\author[bologna]{N.~Le~Neindre},
\author[ts]{G.~V.~Margagliotti},
\author[mi]{A.~Moroni,},
\author[bologna]{G.~Vannini}.

\address[bologna]{Dipartimento di Fisica and INFN, Bologna, Italy}
\address[lpc]{LPC Caen (IN2P3-CNRS/ISMRA) et Universit\'{e}
F-14050 Caen C\'{e}dex, France}
\address[GANIL]{GANIL (DSM-CEA/IN2P3-CNRS), B.P.5027, F-14021 
Caen C\'{e}dex, France}
\address[lnl]{ INFN Laboratorio Nazionale di Legnaro, Italy}
\address[mi]{Dipartimento di Fisica and INFN, Milano, Italy}
\address[ts]{Dipartimento di Fisica and INFN, Trieste, Italy}

\maketitle
\begin{abstract}
A global protocol for the thermostatistical analysis of hot nuclear
sources is discussed. Within our method of minimization of variances
we show that the abnormal kinetic energy fluctuation signal recently
reported in different experimental data~\cite{plb,remi}
is a genuine signal of a first order phase transition in a finite system.

\end{abstract}
\end{frontmatter}
%

\section{Introduction}

The existence of systems exhibiting a negative heat capacity has been
postulated in the seventies in the context of collapsing self-gravitating
systems~\cite{gravity}. The physical origin of the apparent paradox of an
object cooling while absorbing thermal energy is the impossibility of
defining a thermodynamical limit for systems interacting via non
saturating forces. A similar situation occurs in mesoscopic systems where
the range of the interaction, though short, is comparable to the linear
dimension of the system~\cite{gross}. In all these situations the appearance
of a negative heat capacity branch for isolated microcanonical systems is
predicted by theory as a particular example of the specific character of
first order phase transitions in finite systems~\cite{prl99}.

From the experimental point of view, the first evidences of a negative heat
capacity have been reported very recently~\cite{plb,remi,schmidt}. In the
Haberland experiment~\cite{schmidt} small sodium clusters have been
demonstrated to show a negative heat capacity in the region of the solid to
liquid transition, through the measurement of curvature anomalies in the
energy distribution near the transition temperature. In the
case of the liquid to gas like phase transition of hot nuclear 
sources~\cite{plb,remi} negative heat capacities have been observed via 
the study of 
kinetic energy fluctuations~\cite{analytical}. 
In the MULTICS-MINIBALL experiment an
isotropic source of approximately 200 particles has been selected in
peripheral and semi-peripheral collisions of Au nuclei impinging on a Au
target at 35 MeV per nucleon~\cite{plb,michela}. In the INDRA experiment
quasi-fusion isotropic sources of approximately the same size have been
extracted in {\it Xe+Sn} collisions with a beam energy varying from 32 to 50 MeV
per nucleon~\cite{remi,xesn}. The two data samples involving different
reaction mechanisms and measured with different experimental devices give a
negative heat capacity signal in the same excitation energy range. These
measurements not only represent a very strong evidence of the expected
liquid to gas phase transition of nuclear matter~\cite{richertrep}, but also 
show that nuclear thermodynamics can start to be addressed in a quantitative 
way.

A clear evaluation of the degree of reliability of these results is
therefore of a prime interest for the understanding of the nuclear equation
of state as well as the general topic of phase transitions in finite
systems. This is not trivial, since the measurement of kinetic energy
fluctuations is confronted with many technical as well as conceptual
difficulties. 

First, a very careful analysis has to be performed in order to
select an equilibrated or close to equilibrium source from the highly
dynamical process of heavy ion collisions. 

Then, in order to perform the
fluctuation analysis, the energy deposited in the system should be fully
measured on an event by event basis. This ideal situation is never realized
in an actual experiment and one has to check that the limitations and
inefficiencies of the experimental devices do not distort the expected
signal. 

Finally secondary evaporation has to be deconvoluted from the
asymptotically detected partitions, in order to study the correct energy
balance at freeze out. This can only be done via working hypotheses that
have to be carefully scrutinized and eventually constrained by experimental
results.

In this paper we want to analyze in detail all the possible sources of
uncertainty in the heat capacity analysis, coming from the missing
information of data. We wish to establish a general protocol for a
thermostatistical analysis of nuclear sources that allows to deal with mean
values as well as higher order moments of any observable within a minimum
bias technique. The analysis will here be specialized to heat capacity
measurements, but the problems we will be confronted with are common to all
high order moment analysis of multiparametric data~\cite{botet}. We will
demonstrate that within this general protocol the negative heat capacity
signal cannot be artificially generated by the bias of the analysis. We
show that the location of the coexistence zone in the nuclear phase
diagram and the value of the latent heat can be approximately evaluated but
they are still subject to large uncertainties due to the limitations of the
present detection devices.

The plan of the paper is as follows: in section 2 the principles of the 
fluctuation analysis and the present experimental status are summarized. 
In section 3 the fluctuation technique is applied to some well known models 
of nuclear fragmentation. Section 4 briefly addresses the problem of the correct
sorting variables to perform a thermostatistical analysis and the effect of
binning on observables. In section 5 the general method of dealing with a 
missing information in order to minimize the bias of the analysis is presented. 
This method is specialized in the two following sections to the
reconstruction of primary partitions (section 6) and to the distortions
induced by the imperfect calorimetry (section 7). In section 8 the physical
parameters entering the analysis are discussed in detail and some
methods to constrain the value of these parameters with information coming
from the experimental data are proposed. 
Finally section 9 contains conclusions and outlooks.


\section{Kinetic energy fluctuations and microcanonical heat capacity}

First order phase transitions in finite systems are univocally defined by
the abnormal convexity of the thermodynamical potential~\cite{gross,prl99}
in the state variables plane. 
For a microcanonical system undergoing a liquid gas phase
transition, this anomaly produces a characteristic behaviour of the heat
capacity: a negative branch delimited by two divergences that define the
crossing of the coexistence line. This behaviour is suppressed only if sharp
boundary conditions are imposed on the system, i.e. if volume can be
considered as the relevant state variable of the microcanonical statistical
ensemble~\cite{richert}. This is certainly not the case for nuclear sources
which are open systems not constrained by any boundary condition. In the
experimental situation the volume is rather an observable known at best in
average and impossible to use as a sorting variable. In such a case the
pressure, interpreted as the Lagrange multiplier associated to the volume,
appears to be the relevant state variable together with the total deposited
energy; a convex intruder is expected and the energy fluctuations (see
below) are related to $C_{p}$.

In Ref.~\cite{analytical} it was proposed that the microcanonical heat
capacity can be measured using partial energy fluctuations. For a classical
system with momentum independent interactions the total energy $E$ can be
decomposed into two independent components, its kinetic ($E_{k}$) and
interaction energy ($E_{I}$). Since the energy partition directly depends on
the partial entropies $S_{k}$ and $S_{I}$, the kinetic energy variance can
be related, in the Gaussian approximation, to the heat 
capacities~\cite{analytical,lebowitz}: 
\begin{equation}
A_{0} \sigma_{k}^{2}\simeq {T}^{2}\frac{c_{k} c_{I}}{c_{k}+c_{I}}  \label{eq:sigma}
\end{equation}
where $c_{k}$ and $c_{I}$ are the kinetic and interaction microcanonical
heat capacities per particle calculated for the most probable energy partition
characterized by a microcanonical temperature ${T}$. Equation (\ref{eq:sigma}) 
can be inverted to extract, from the observed fluctuations, an estimate of
the heat capacity~\cite{analytical}: 
\begin{equation}
\frac{C}{A_{0}} = c \simeq c_{k}+c_{I}\simeq \frac{c_{k}^{2}}{c_{k}-A_{0}\sigma _{k}^{2}/{T}^{2}}
\label{eqcentral}
\end{equation}
From eq.(\ref{eqcentral}) we can see that the heat capacity
becomes negative if the kinetic energy fluctuations overcome the canonical expectation 
$A_{0}\sigma _{k}^{2}/{T}^{2}=c_{k}$. It is amazing that the constraint of energy
conservation leads in the phase transition region to larger fluctuations
than in the canonical case where the total energy is free to fluctuate. This
is because the kinetic energy part is forced to share the total available
energy with the interaction part. When the interaction part presents a
negative heat capacity, the jumps from liquid to gas induce strong
fluctuations in the energy partitioning. It is also interesting to remark
that if the kinetic equation of state is known, the most probable value of
the kinetic energy, as well as the average one, acts as a very powerful
microcanonical thermometer~\cite{analytical}.

In principle one could argue that the same information on the heat capacity
can be obtained by taking the derivative of the correlation between the
temperature and the excitation energy (the so called caloric curve). We want
to stress that fluctuations are characteristic of the state and so depend on
the pertinent state variable while the caloric curve $T(E)$ depends upon the
specific thermodynamical transformation from one state to another.
Therefore, the information obtained by taking the derivative of the measured
caloric curve may differ from the information coming from the 
fluctuations~\cite{iso_iso}.

In order to perform a thermostatistical analysis one has to collect a data
sample which corresponds to a (collection of) microcanonical ensembles. The
microcanonical ensemble is relevant for the analysis of experimental data
because of the absence of a heat bath and since using calorimetry techniques
the excitation energy can be measured on an event-by-event basis. For any
arbitrary shape of the excitation energy distribution the events can thus be
sorted in constant energy bins, i.e. in microcanonical ensembles.

Single source complete events have to be selected with a constant value for
the collected charge in each energy bin. In both analyses only well detected
events (total detected charge larger than 70\%~\cite{plb} or 80\%~\cite{remi}
of the Au charge) were considered. After a shape analysis~\cite{cugnon}
central collisions are isolated from the {\it Xe+Sn} collisions with a selection
on the flow angle $\Theta_F$ between the beam axis and the main eigenvector
of the kinetic energy tensor ($\Theta_F >60^o$). For the {\it Au+Au} system,
peripheral collisions of a predominantly binary character are selected by
requiring the velocity of the largest fragment in each event to be at least
75\% of the beam velocity. Then fragments of each event are considered as
originated by the quasi-projectile if forward emitted in the ellipsoid
reference frame. Moreover in both data sets only events where the total
reconstructed source charge results within 10\% of the $Au$ charge are kept.
For details about the selection criteria, see \cite{plb,remi} and the
references quoted therein. For both reactions the selected events are close
to the maximum of the total charge distribution. As it will explained below 
in more
detail, this means that source mass fluctuations are under control and that
we are dealing with a statistically significant sample of each centrality
(i.e. excitation energy) bin.

\begin{figure}[htb]
\begin{center}
\epsfig{figure=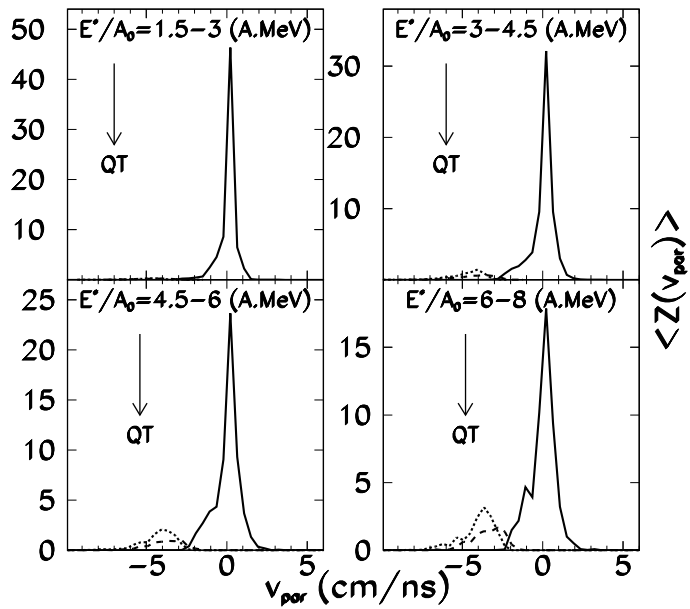,width=10.cm}
\end{center}
\caption{\it Detected average fragment charge ($Z \ge 3$) for the peripheral 
{\it Au+Au} collisions as a function of the parallel velocity in the 
quasi-projectile
reference frame, for different excitation energy bins. Full lines: fragments
belonging to the quasi-projectile. Dashed lines: fragments rejected from the
quasi-projectile selection. Dotted lines: simulated and filtered quasi-target
contribution.}
\label{fig:1}
\end{figure}

To check the quality of the source selection criteria, a standard procedure
consists in verifying that events are spherically symmetric in momentum
space. Of course it is very likely that in the dynamical preparation of
the nuclear source the shape degree of freedom may not be completely
relaxed; such a case could in principle be addressed within a statistical
ensemble where deformation is explicitly accounted for. However restricting
the analysis to spherical systems guarantees that non statistical effects,
as preequilibrium or midrapidity emission, do not pollute the statistical
sample.

In the case of central collisions the isotropic character of fragment emission is
implicit in the flow angle selection and has been verified in detail
comparing to simulations~\cite{xesn,frankland}. Concerning the
quasi-projectile (QP) events, the charge density~~\cite{lecolley} 
of selected events is plotted in Figure 1 in different excitation energy bins. The
QP fragments are normally distributed even for the less peripheral
collisions. Moreover the backward emitted fragments in the ellipsoid
reference frame (which are excluded in the subsequent analysis) are
perfectly consistent with the filtered simulation of a quasi-target source
symmetric to the QP. This suggests that midrapidity emission of fragments is
negligible for the selected events (see Fig.s 1~f) and 3 of Ref.~\cite
{michela}), at variance with other experimental studies~\cite{msuganil},
where different entrance channels or selection criteria enhance a dynamical
emission at semi-peripheral impact parameters.

More delicate is the contribution of light particles ($Z= 1, 2$)
which are likely to be
emitted during the whole collision process and not confined to the freeze
out stage. For the {\it Xe+Sn} central collisions, in order to avoid the 
possible
contamination of preequilibrium, only particles emitted between $60^o$ and 
$120^o$ in the center of mass are considered and their contribution is
doubled to compensate for the forward/backward anisotropy~\cite{remi}. In
the same way for the Au quasi-projectile, a possible contribution of
midrapidity emission is minimized by substituting the backward light
particle emission by the symmetric of the forward emission, in the
quasi-projectile reference frame~\cite{plb}.

The excitation energy of the source is reconstructed from calorimetry on an
event by event basis~\cite{michela,xesn}. Different hypotheses have been
considered about the number and energy of the undetected 
neutrons~\cite{michela} and will be discussed in detail in the following.

The relevance of the thermostatistical analysis depends on the approximation
at which an equilibrium is realized. Equilibrium, being by definition the
reducibility of the multidimensional information to a number of macroscopic
constraints, can never be proved but only be disproved, even for an
isotropic source. As a general statement, the degree of
approximation of an equilibrium is indicated by the degree of the agreement
of data with the prediction of a statistical model containing the same
constraints as the data. For both data sets discussed in this 
paper~\cite{remi,michela,xesn,prctrieste} a very good agreement has been 
found with
the predictions of the statistical model SMM~\cite{bondorf} for static as
well as dynamical variables. In particular a detailed analysis of the
average kinetic energy of the fragments indicates the presence of a
collective component ranging from 0 A.MeV for the most peripheral collisions
of the {\it Au+Au} system to 2 A.MeV for the central {\it Xe+Sn} ones at 
50 A.MeV incident energy. The good model reproduction of several 
observables suggests
that, in this energy range, the collective motion can be to a good
approximation superimposed to a thermal picture. The collective contribution
to the excitation energy has been removed in the subsequent analysis.

Let us now turn to the heat capacity measurement. In order to extract the
heat capacity from eq.~(\ref{eqcentral}) one has to decompose the total
measured excitation energy into the kinetic and interaction contributions at
the time of fragment formation, i.e. at freeze out. In the case of nuclear
fragmentation data, this is complicated by the fact that the fragments are
detected at infinity, after secondary de-excitation, $i.e.$ with lower
masses. Moreover, because of the presence of the long range Coulomb
interaction, asymptotic kinetic energies have to be corrected for the
Coulomb boost. To take into account these distortions, the kinetic energy at
freeze out is reconstructed by applying the energy balance event by event: 
\begin{equation}
E_k = m_0 + E^* - \sum_{i=1}^M m_i - E_{coul}(V_{FO}) = E^* - E_I
\label{eq3a}
\end{equation}
where $m_0$, $m_i$ are the mass excess of the source and of the primary
products respectively, $E^*$ the excitation energy measured via calorimetry,
and $E_{coul}$ is the Coulomb energy of the partition. The two important
unknown quantities here are the primary multiplicity $M$ entering the 
$Q$-value and the freeze-out volume ($V_{FO}$) determining the Coulomb energy.
Only qualitative information about these quantities are given by theory. 
Dynamical models predict a sudden increase of density fluctuations leading
to fragment formation at low density, but the actual value of the volume
depends on the model. Concerning secondary cooling, a correct description of
the temperature dependent partition function is needed to reconstruct the
de-excitation chain and discriminate between prompt and secondary emission.
At low excitation energies this aim is accomplished by the theory of
compound nucleus. At higher excitation energies the opening of
multifragmentation channels makes an exact evaluation increasingly
difficult. Sophisticated Monte-Carlo calculations of the multifragmentation
pattern are available since many years~\cite{gross,bondorf,elliot}, but the
internal partition function at high temperature is poorly known, and the
suppression procedure of the state-density integral at high excitation
energy is not unique.

The Coulomb energy has been evaluated in Ref.s~\cite{plb,remi} by randomly
positioning non overlapping spherical primary products in a freeze-out
volume from 3 to 6 times larger than the normal volume. The primary
multiplicity and the primary masses have been obtained~\cite{michela} by
sharing the final light particles and neutrons among the detected fragments,
following two extreme freeze-out hypotheses~\cite{gross,bondorf}. A lower
limit for the primary multiplicity can be obtained by assuming ({\it hot
fragment} hypothesis) that the totality of light charged particles and
neutrons is emitted by the hot primary fragments in a secondary decay
process. An upper limit ({\it cold fragment} hypothesis) consists in
assuming that all light charged particles are primary and de-excitation
concerns only neutron evaporation. 

To apply eq.~(\ref{eqcentral}) we also need to calculate the microcanonical
temperature $T$ of the system. An estimator of $T$ can be obtained by
inverting the kinetic equation of state 
\begin{equation}
\langle E_{k}\rangle =\big\langle \sum_{i=1}^{M}a_{i}\big\rangle T^{2}
+\big\langle \frac{3}{2}(M-1)\big\rangle T  \label{temp}
\end{equation}
where the brackets $<~^{.}~>$ indicate the average on the events with the
same $E^{*}$ and $a_{i}$ is the level density parameter~\cite{michela}. 
The temperatures of the system obtained
in the two freeze-out hypotheses can be considered as an upper and a lower 
limit of the actual freeze-out temperature~\cite{michela}.
\begin{figure}[tbh]
\begin{center}
\epsfig{figure=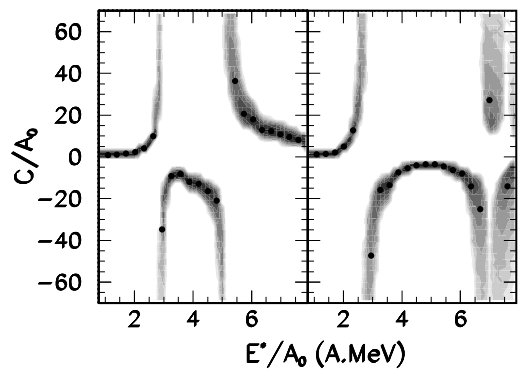,width=10.cm}
\end{center}
\caption{\it Heat capacity per nucleon (solid symbols) obtained from 
equation \protect({\ref{eqcentral}}) for the QP data. The panel on the left 
corresponds to hot
primary fragment in a freeze-out volume $3\ V_0$. The panel on the right
corresponds to cold primary fragments in a freeze-out volume $6\ V_0$. The
grey contour indicates the confidence region for $C / A_0$.
} 
\label{fig:2}
\end{figure}

The functional form of eq.(\ref{temp}) is certainly a reasonable ansatz
for the average kinetic energy of an ensemble of fragments, but an extra source
of uncertainty comes from the value chosen for the level density parameter.
This will be discussed in Section 8.

Figs. 2 and 3 show the final result for the total heat capacity obtained
from the Au quasi-projectile and the central {\it Xe+Sn} 
collisions~\cite{plb,remi}.
\begin{figure}[tbh]
\begin{center}
\epsfig{figure=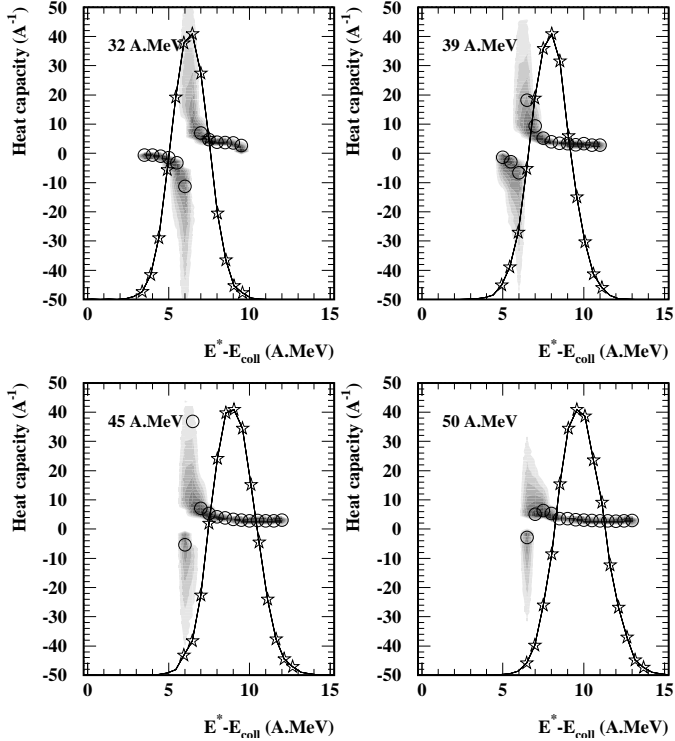,width=10.cm} 
\label{fig:3}
\end{center}
\caption{ \it Specific heat capacity (open points) obtained from 
eq. \protect({\ref{eqcentral}}) for the central Xe+Sn data at four 
different bombarding
energies within the hot fragment hypothesis as calculated in 
ref.~\protect{\cite{remi}}.
Lines and stars: calorimetric excitation energy distributions with collective
flow correction. The grey contour indicates the confidence region for $C / A_0$.}
\end{figure}
$c_{k}$ in eq.(\ref{eqcentral}) has been obtained by taking the numerical
derivative of $\langle E_{k}\rangle $ with respect to $T$. The grey contour
represents the $C$ distribution, evaluated through a Monte-Carlo error
simulation program. Only statistical errors are taken into account in the
evaluation of this confidence region. A distinct negative branch appears
pointing to a 1-st order liquid-gas phase transition, the distance between
the poles being associated with the latent heat. A wide range of 
impact parameters, leading to a widely spread excitation function, is
available for the quasi-projectile data, allowing to follow the whole
behavior of the heat capacity in the pure as well as in the mixed phases. On
the other side in the central collisions sample the selection of the
quasi-fused source isolates very central impact parameters and the heat
capacity can be measured only around the transition on the vapor side. In
this energy interval the QP data are subject to some uncertainties due to
the lack of statistics and the difficulties in the source selection, however
the agreement between the two sets of data is remarkable. A negative heat
capacity branch appears in the same excitation energy range in these
different collisions leading to different reaction mechanisms, detected with
different experimental devices, selected and analyzed with independent
methods. 

A word of caution is however necessary. 
Negative heat capacity is obtained when kinetic energy fluctuations exceed
the value of the kinetic heat capacity $c_k$ (see eq.(\ref{eqcentral})). $c_k$ 
itself has an upper bound given by the classical Boltzmann limit $c_k \leq
3/2$. One may wonder if the inefficiencies of the apparatus and the
reconstruction hypotheses may introduce an uncontrolled source of
fluctuations that may mock up the negative heat capacity signal. In the
following sections we will address in great detail all the
possible sources of uncertainty related to the fluctuation analysis.


\section{Can we measure a positive heat capacity?}

A first global check of the experimental method consists in applying the
fluctuation analysis to a sample of theoretical events generated with
statistical models. Two well known and sophisticated models, often used to
simulate heavy ion reaction data, are SMM~\cite{bondorf} and 
GEMINI~\cite{charity}. 
\begin{figure}[htb]
\begin{center}
\epsfig{figure=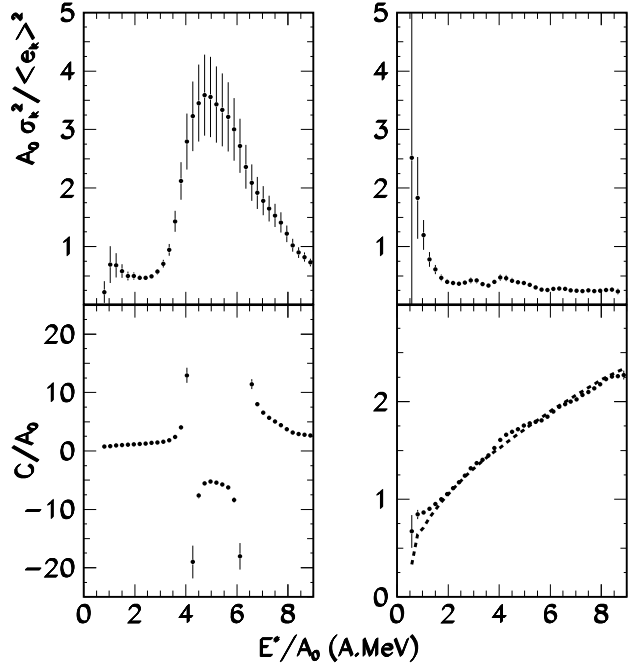,width=10.cm}
\end{center}
\caption{\it Normalized kinetic energy fluctuations (upper part) and the
corresponding reconstructed heat capacity (lower part) for SMM (left) and
GEMINI (right) simulations as a function of the excitation energy. Dashed
line: numerical derivative of the theoretical caloric curve.}
\label{fig:4}
\end{figure}
SMM aims to describe multifragmentation at low freeze out density and for
the typical volumes used (around three times the normal source volume) shows
a first order liquid-gas like phase transition~\cite{alex}. GEMINI describes
fragment production as a sequence of binary fission-like emissions at a
density close to the normal nuclear matter density. The thermodynamics of
this model has, to our knowledge, never been scrutinized. Because of the
hypothesis of low emission rate (i.e. low vapor-like pressure) and high
density we expect that the model should be close for all excitation
energies to the coexistence line on the liquid border, and it should not
show negative heat capacity. We notice incidentally that the data we are
discussing are very well reproduced by SMM at all excitation
energies~\cite{michela,xesn} while the GEMINI code can only reproduce the
most peripheral QP data.

The procedure sketched in the previous section has been applied to simulated
SMM and GEMINI data in the energy range between 1 and 9 A.MeV. In both cases
the freeze out has been reconstructed within the {\it hot fragment}
hypothesis~\cite{plb}. Figure 4 shows the resulting kinetic energy
fluctuations as well as the corresponding heat capacity. The
normalized fluctuations show a peak in the SMM data while they are
monotonically decreasing in the case of the GEMINI model. This tends to
confirm the claim~\cite{analytical} that a peak in the reduced kinetic
energy variance is by itself a signature of a thermodynamical phase
transition. If in addition the temperature is estimated via eq.(\ref{temp}),
it is possible to convert the fluctuation signal into a heat capacity.
Abnormal fluctuations are obtained for the SMM simulation while the heat
capacity for GEMINI is a monotonically increasing function of the excitation
energy. It is interesting to remark that for the GEMINI calculation the
reconstructed heat capacity is very close to the derivative of the caloric
curve (dashed line in figure 4) even if the freeze out reconstruction is
certainly not suited to a sequential decay scenario. This can be understood
since in a sequential scenario a relatively low number of channels is open.
The fluctuations of the interaction energy $E_{I}$ are then very small and
eq.(\ref{eqcentral}) gives a heat capacity close to the kinetic heat
capacity. For the same reason the functional dependence on temperature of
the reconstructed kinetic energy is similar to the total caloric curve. In
addition eq.(\ref{temp}) appears to give a reasonable estimate of the
temperature of the model, the integration over the deexcitation chain being
only a minor correction. This result seems to indicate that eq.(\ref
{eqcentral}) is a powerful tool to extract heat capacities in presence or in
absence of a phase transition, for sequential as well as simultaneous
emission processes.


\section{The effect of binning}

A very first concern when applying a microcanonical analysis to a set of
(experimental or simulated) data is the effect of event mixing due to
binning. Equation (\ref{eqcentral}) is derived under the microcanonical
constraint, i.e. a strict conservation of energy and mass. The process of
energy binning violates energy conservation while the dynamical fluctuations
of the entrance channel and the lack of detection of neutrons and fragment
masses induce fluctuations in the source size in each excitation energy bin,
which is controlled at the 10-20\% level only~\cite{plb,remi}.

The effect of varying the width of the energy bins is explored in the left
part of figure 5 using a SMM simulation. 70000 events are generated from a
continuous and flat distribution of excitation energies in the range 1-10
A.MeV for an Au source. This statistical sample will be also used for all
the successive analysis presented in this paper. In all cases the freeze out
is reconstructed, within the {\it hot fragments hypothesis}, from the 
simulated final partitions as in the experimental data (see section 2). 
To isolate the genuine effect of binning, the exact
input excitation energy of the model is used in each event. Additional 
distortions due to the deficiencies of the experimental calorimetry will be
discussed in section 7.

From figure 5 one can see that the average values of the presented
observables are too smoothly varying to be affected even by a non realistic
binning as wide as $\Delta E = 0.9$ A.MeV. When the width of the bin is
increased, fluctuations are smoothed out but the height of the maximum is
not affected. This is easy to understand since any (small) violation of the
energy conservation perturbs the microcanonical constraint $E^*=E_k+E_I$.
This flattens the normalized fluctuation $\sigma_k^2/\langle e_k \rangle^2$
similarly to the fluctuation suppression operated by the microcanonical
event mixing in the canonical ensemble. The effect is barely visible for bin
widths smaller than about 1 A.MeV. A similar effect is
obtained if a (fluctuating) non thermal component is added to the excitation
energy~\cite{remi} to simulate the effect of an imperfect subtraction of the
collective flow in central collisions. The same analysis applied to the data
shows no sizeable effect due to the bin width~\cite{bormio_michela}. The
fluctuation smoothing is barely visible because the calorimetric
uncertainties cannot be disentangled from the effect of binning (see next
sections). 
\begin{figure}[htb]
\begin{center}
\epsfig{figure=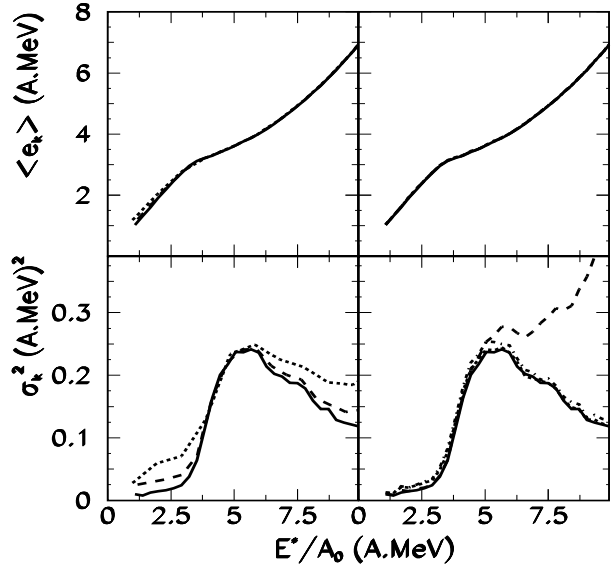,height=9.cm,width=10.cm}
\end{center}
\caption{\it First (upper part) and second moment (lower part) of the SMM
kinetic energy distribution reconstructed within the hot fragment 
hypothesis~{\protect\cite{plb}}. Left part: different energy bin widths $\Delta
(E^*/A_0) = 0.3$ (full line), $0.6$ (dashed line), $0.9 A.MeV$ (dotted line). 
Right part: effect of the non conservation
of mass induced by the doubling of light particles (see text).}
\label{fig:5}
\end{figure}

Next we turn to the possible spurious fluctuations induced by violations of
mass conservation. As it is done with the data, only the light particles
emitted in one half of the solid angle in the SMM simulation are kept and
their contribution is symmetrized backward. Once again we leave the effect
of this doubling procedure on the calorimetric reconstruction of excitation
energy to section 7, and use here the input energy of the model. The
doubling of light particles leads to mass number fluctuations that increase
with the excitation energy and that are comparable to the experimental mass
width in each excitation energy bin. The right part of figure 5 shows that
this mass fluctuation does not affect average quantities while a strong
deformation of the original fluctuation signal (full line) is apparent if we
calculate the variance of the total kinetic energy (dashed line). If
energies per nucleon are used (dash-dotted line) the signal is almost not
perturbed. If events of mass number differing more than 10\% from the
original source mass are rejected (dotted line) the initial fluctuations are
also approximately restored. Both procedures are used for the analysis of
the data. Since these general results do not depend on the details of the
simulation employed, we can be confident that no spurious fluctuations are
generated either by the procedure of light particles doubling or by the 10\%
fluctuations of the source size.


\section{Dealing with missing information}

Deriving nuclear thermodynamics from the fragmentation sample on an event by 
event basis, one is systematically confronted with the problem of missing 
information. This concerns detection
limitations (neither neutrons nor fragment masses are measured, the response
of the experimental filter deforms the events) as well as the uncertainties
of the freeze out reconstruction (the primary masses, multiplicities and
freeze out volume are not known) and the unknown physical parameters (the
kinetic equation of state eq.(\ref{temp}) is only an ansatz, and in
particular the level density parameter is poorly known at high excitation
energy). Since we are interested in first as well as second order moments,
the missing information has to be implemented event by event. In principle
one would like to restore the missing information in such a way that mean
values as well as variances agree with values obtained for the same
observables from independent measurements. This is however in general a very
tough task. As an example one may be able to measure an average value for
the size of the fragmenting source through correlation techniques, but this
method does not give an event by event response. Only sophisticated
backtracing procedures~\cite{pierre}, working in the multivariate space of
the source characteristics, provide event by event estimates. However, the
obtained distributions turn out fully model dependent.

Here we propose a less ambitious way of dealing with missing information.
The heat capacity measurement results from the simultaneous evaluation of
partial energy $e_{k}$ fluctuations, total deposited energy and temperature.
Heat capacity is negative if the variance $A_{0}\sigma _{k}^{2}$ exceeds a value 
$T^{2}c_{k}=T^{2}\frac{d<e_{k}>}{dT}$ determined solely by the same partial
energy mean value $<e_{k}>$ through its equation of state $T(e_{k})$. Let us
suppose that the mean value of the missing observables is known from theory
or from an independent measurement. Then the event by event value of the
same observables can be fixed to its mean value, such as to systematically
minimize the partial energy fluctuations. If then the negative heat capacity
signal survives, this cannot be attributed to spurious fluctuation due to not
measured quantities. It is important to stress that this conservative
attitude guarantees the physical meaning of abnormal partial energy
fluctuations but prevents a quantitative analysis of the phase transition,
in particular a precise evaluation of the latent heat.

The partial energy $e_{k}$ (i.e. $E_{k}/A_{0}$) is obtained as 
the difference between the total
deposited calorimetric energy and the interaction energy at freeze out 
(eq.(\ref{eq3a})). The general conservative philosophy of suppressing
fluctuations of not measured quantities is applied in the next section to
the missing observables that enter in the interaction energy, and in section
7 to the calorimetry. The average values of not measured observables will be
then fixed from independent experimental constraints in section 8.


\section{The freeze out reconstruction}

Let us take again SMM as an event generator. The freeze out temperature and
kinetic energy variance of the model are shown as thick solid lines in
figure 6. In this calculation a constant volume $V=3V_{0}$ is used. The
advantage of dealing with simulated data is that we can study the effect of
the interaction energy reconstruction on $E_{k}$, independent of the
calorimetry errors. The excitation energy in figure 6 is the input
excitation energy of the model. In such a situation the only parameters left
in eq.(\ref{eq3a}) are the freeze out observables, namely the volume
entering the Coulomb energy and the primary masses and multiplicity entering
the Coulomb term as well as the Q-value. We can take the asymptotic
simulated partitions and fix on average these parameters such as to
reproduce the theoretical average kinetic energy at freeze out. The way this
average information is implemented event by event implies an enhancement or
a suppression of the second moment. This can be predicted a priori and does
not depend on the specific model that generates the events, used in this
paper only for illustration.

\begin{figure}[htb]
\begin{center}
\epsfig{figure=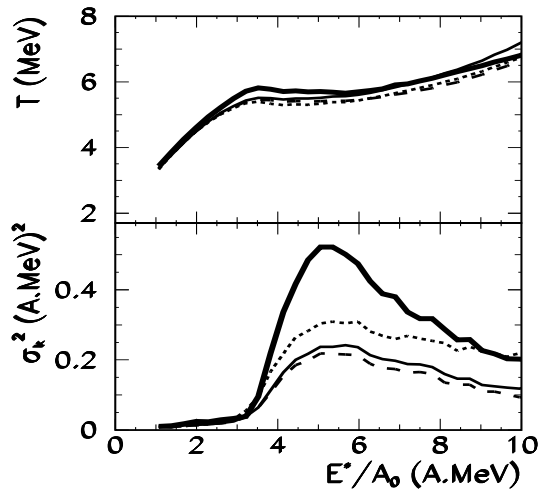,height=9.cm,width=9.cm}
\end{center}
\caption{\it Temperature (upper part) and kinetic energy variance (lower part)
of SMM (thick solid lines) compared to the estimate from the different
freeze out reconstructions (see text).}
\label{fig:6}
\end{figure}

The thin solid line corresponds to the {\it hot fragment} 
hypothesis~\cite{plb}. All light charged particles and neutrons in each event 
are shared among the fragments such that they get the same $N/Z$ of the source. 
(The same result is obtained if, to evaluate the masses of primary fragments, we
use a functional $A(Z,E^*)$ fitted to SMM primary fragments). The average
freeze out multiplicity is thus minimized. Since a deterministic algorithm
is used to construct the freeze out multiplicity, the event by event
difference between the asymptotic and freeze out multiplicities is
suppressed. Not surprisingly, the resulting interaction energy variance
strongly decreases. This schematic reconstruction hypothesis was used on
experimental data to represent a lower limit to the average freeze-out
multiplicity.

In the framework of SMM, this hypothesis is not very realistic, a number of
light charged particles being present at the freeze out stage. If we allow
an excitation energy dependent percentage of primary light particles (dashed
lines), the mean value is better reproduced especially at high energies, but
the variance is not affected. Whatever deterministic algorithm we can choose
to reconstruct the primary multiplicity, it will imply a suppression of
fluctuations if it is implemented on an event by event basis. This is due to
the fact that the multiplicity is positively correlated both to the Coulomb
potential $V_c$ and to the $Q$-value; a reduction in the multiplicity
fluctuation will therefore suppress the variance of $V_c$ and $Q$ as well as
their (positive) covariance. This is why the increase of $\sigma_k$ is
negligible, even if we impose a large average value of the multiplicity
(within the constraint of an approximate reproduction of $<e_k>$).

On the other hand, any hypothesis that assigns to a missing information a
value fluctuating event by event has the effect of increasing $\sigma_k$. As
an example, the dotted lines in figure 6 are obtained by assuming a flat
distribution of freeze-out volumes with a width equal to the average value.
The (moderate) increase of $\sigma_k$ indicates that, if in the
fragmentation data the freeze out volume does fluctuate event by 
event~\cite{iso_iso}, our analysis, which assumes a constant volume in each 
excitation energy bin, once again underestimates the physical fluctuations. 
In all the cases depicted in figure 6 the same level density parameter as in 
the model~\cite{lev_dens} has been used; as a consequence the degree of 
reproduction of $<e_k>$ and $T$ is comparable.


\section{The uncertainties induced by calorimetry}

In the experimental evaluation of the kinetic energy at freeze out (eq.(\ref
{eq3a})) the total excitation energy is not a fixed external parameter as we
have assumed in the previous sections, but it is the result of an event by
event calorimetric measurement via 
\begin{equation}
E^*=\sum_{i=1}^{N_c} \left( m_{i} + E_{i}\right) 
+ N_n \left ( m_n + <E_{n}> \right) - m_0
\end{equation}
Here $N_c$ and $N_n$ are the charged particles and neutron multiplicities, 
$E_{i}$ ($E_{n}$) are the kinetic energies, $m_{i}$ ($m_n$) are the mass 
excesses of
the charged reaction products (neutrons) and $m_0$ is the mass excess of the
source. The calorimetric measurement is affected by a number of
uncertainties. The neutron number as well as the neutron energies are not
measured; an isotopic resolution is achieved only for light fragments; the
experimental filter may deform the energy response though this effect is
minimized restricting to only quasi complete events (see
section 2). Moreover for the analysis presented here only the light
particles emitted in one half of the total solid angle are kept and their
contribution is symmetrized to the other half.

To complete this missing information a mass has been assumed for all the
detected fragments following the EPAX~\cite{epax} parameterization; the
number of neutrons is then deduced from mass conservation by assuming that
the fragmenting source has the same isospin ratio as the composite system
(for central events) or as the projectile (for quasi-projectile events). The
average neutron energy has been obtained for each excitation energy bin
from the total detected kinetic energy, by means of an effective 
temperature~\cite{eos}.

\begin{figure}[htb]
\begin{center}
\epsfig{figure=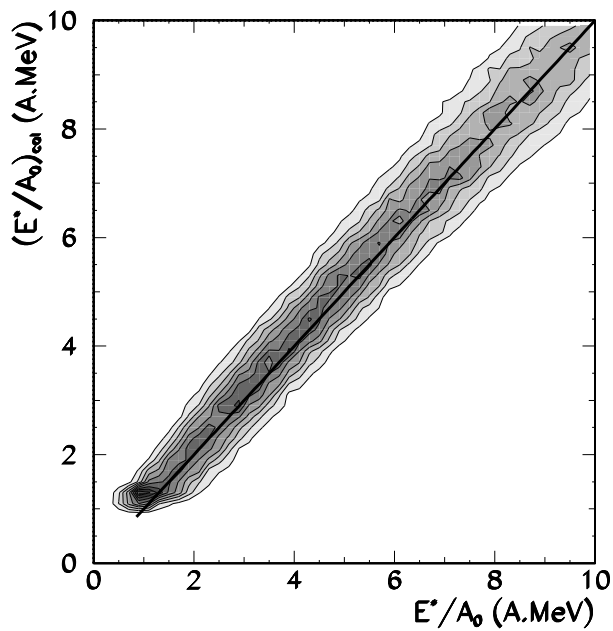,width=10.cm}
\end{center}
\caption{\it Correlation between the calorimetric measurement of the excitation
energy per nucleon and the input energy of SMM simulations.}
\label{fig:7}
\end{figure}

The effect of all these approximations is displayed in figure 7, which shows
the correlation between the calorimetric excitation energy, calculated with
the same uncertainties present in the data, and the input excitation energy
of SMM. The average value is reproduced within 0.5 A.MeV at the highest
excitation energies. The average difference does not exceed 0.2 A.MeV 
in the negative heat capacity region. This fixes the minimum bin width that 
can be used for the data analysis. However an important dispersion around the 
most probable value is visible. 
It is important to stress that this width plays a very different role with
respect to the freeze out reconstruction uncertainties discussed in the
preceding section. In fact, data being analyzed in constant excitation
energy bins, the variance of the excitation energy reconstruction does not
sum up with the interaction energy variance in eq.(\ref{eq3a}). The spurious
width induced by calorimetry is determined solely by the energy bin,
independent of the width of the calorimetric excitation energy distribution
of figure 7. In section 4 we have already shown that the effect of the bin
width on $\sigma _{k}$ is not dramatic.

\begin{figure}[htb]
\begin{center}
\epsfig{figure=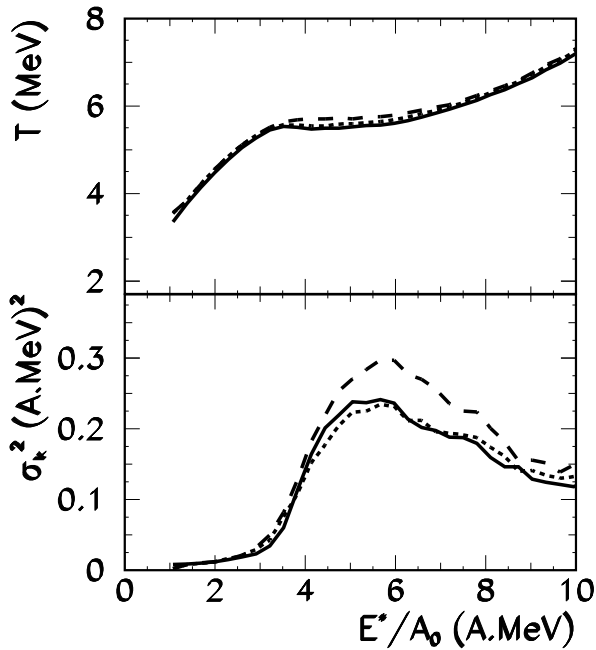,width=10.cm}
\end{center}
\caption{\it Effect of calorimetry on the temperature (upper part) and kinetic
energy variance (lower part) measurement. Solid lines: freeze out
reconstruction as in section 5 and exact excitation energy. Dashed lines:
treatment of data as in the experimental sample. Dotted lines: same as
dashed, but only events within 10\% from the input energy are retained.}
\label{fig:8}
\end{figure}

The calorimetric uncertainties, however, can cause an additional spurious
effect. A very large spread of the measured excitation energy, if used as a
sorting parameter, can lead to event mixing, which can in turn artificially
enhance the interaction energy fluctuations. Once again the relative
importance of event mixing can be estimated only by a simulation.

Figure 8 shows the effect of calorimetry on the two ingredients necessary to
estimate the heat capacity. The solid lines are the same as the thin solid
lines of figure 6. They  give the reference calculation where both $T$ and 
$\sigma _{k}$ are calculated from the event by event reconstructed $E_{k}$
and the excitation energy per nucleon is the input $E^{*}/A_{0}$ of the
simulation. The event mixing due to the imperfect calorimetry (dashed line)
does not affect the calculation of average observables like the temperature.
On the other side the variance is somewhat enhanced and its functional
behavior slightly deformed. If however only events with energy that differs
from the input energy less than 10\% are kept, the reference result is
approximately recovered (dotted lines). Spurious fluctuations can then be
avoided if a constraint is put on the events by means of a conservation law.
This idea has already been exploited in section 3 where the doubling of
light charged particles was shown not to enhance partial energy
fluctuations, if a constraint was put on the size of the reconstructed
source. However, because of the calorimetric uncertainty, the total
deposited energy is not known a priori. In this case the artificial
enhancement of partial energy fluctuations can be minimized if only
''conservative'' hypothesis are applied to the missing information, as we
have already discussed in the previous sections. Every replacement of a not
measured quantity by an estimated average value of the same quantity 
reduces the fluctuations in any observable positively correlated to the
experimentally unknown variable. On the other side, if the missing 
information is replaced with an estimate obtained from measured observables 
of the same event,
this correlation within the same event induces a spurious fluctuation.

\begin{figure}[htb]
\begin{center}
\epsfig{figure=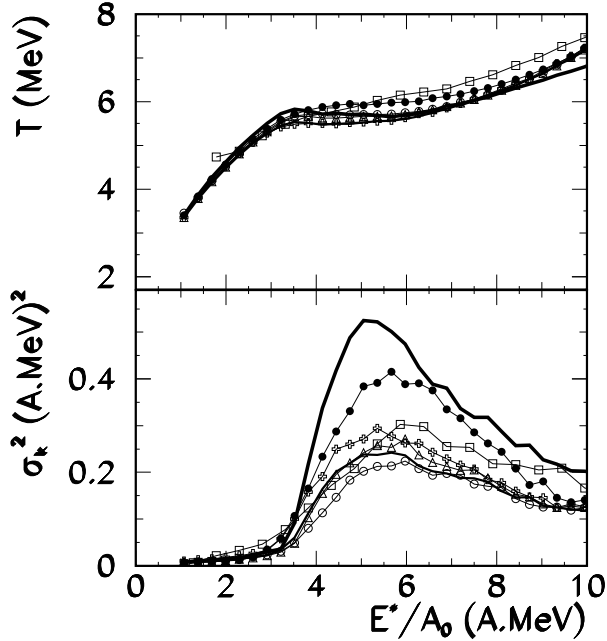,width=10.cm}
\end{center}
\caption{\it Effect of the calorimetry on the temperature (upper part) and
kinetic energy variance (lower part) measurement. Solid lines: as in figure
6. Symbols: effect of the calorimetric hypotheses one by one (see text).}
\label{fig:9}
\end{figure}

This general statement can be better understood by looking at figure 9 which
shows  the deformation  induced by the different unknown quantities one by
one. The thick lines give the exact result of the model, while the thin
solid lines correspond to the hot freeze out reconstruction as in figure 6
and 8. All the other curves show the effect of the different calorimetric
hypotheses which have been applied to the data:

\begin{itemize}
\item  replacing the (fluctuating) neutron energies with the (average)
effective temperature estimate~\cite{eos} (triangles);

\item  replacing the (fluctuating) asymptotic masses with the (average) 
EPAX~\cite{epax} prescription (open circles);

\item  doubling the light charged particles and rejecting events that
violate mass conservation more than 10\% (crosses).
\end{itemize}

Finally the squares in figure 9 correspond to a complete calorimetric
reconstruction as in the data (all the preceding steps summed up), including
also the effect of the MULTICS-MINIBALL fil\-ter. A very similar
distribution is obtained if the INDRA filter is applied. All these steps do
not affect dramatically neither the first nor the second moment of the kinetic 
energy distribution, showing that the calorimetric spread is well
under control. The distortions induced by the filter can be appreciated by
comparing the open squares in figure 9 (which correspond to filtered events)
to the dashed lines of figure 8 (not filtered events). The main effect of
the filter is a general smoothing of the temperature and a slight reduction
of fluctuations in the negative heat capacity region.

The only case that produces an evident distortion (full points in figure 9)
corresponds to a prescription that attributes to neutrons in each event the
same kinetic energy of protons, corrected for the Coulomb barrier. The
correlation between protons and neutrons within the same event causes a
large calorimetric spread leading to a non negligible event mixing. This is
a clear example of an algorithm to be avoided in the analysis of the data. 


\section{The unknown physical parameters: towards a quantitative nuclear
thermodynamics}

In the previous sections we have analyzed the robustness of the fluctuation
signal in a simulation where the average values of the  observables at
freeze-out are known a priori. This is unfortunately not the case for the
experimental situation. One can then wonder how much the observation of
negative heat capacity in the data depends on the values chosen for these
observables. We have already discussed in section 2 the effect of varying
the freeze out multiplicity. Switching between the two extreme
freeze-out hypotheses leaves the position of the first divergence
practically unchanged but strongly modifies the estimated latent
heat (see figure 2). The persistence of the negative heat capacity signal
may look surprising, knowing that the actual values of the temperature as
well as its behavior as a function of the excitation energy are appreciably
different in the two freeze out hypotheses (see figure 8 of 
Ref.~\cite{michela}). 
The fluctuation observable $c$ seems in this sense more robust
than the caloric curve. This is due to the fact that, even if the (model
dependent) temperature in our analysis acts as a normalization factor for
the fluctuations (see eq.(\ref{eqcentral})), the reference fluctuation scale
($c_{k}$) and the fluctuation itself ($\sigma _{k}$) are consistently
derived from the same equation of state. Once the multiplicity is fixed, the
actual freeze-out composition i.e. the precise relationship between mass and
charge of the primary fragments does not modify the results as discussed in
section 6.

\begin{figure}[tbh]
\begin{center}
\epsfig{figure=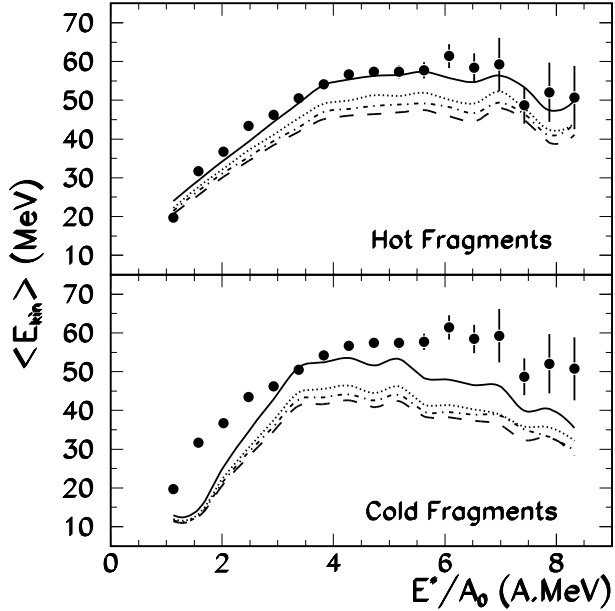,width=10.cm}
\end{center}
\caption{\it Average fragment ($Z\ge 3$) kinetic energy as a function of the
excitation energy for QP data. Symbols: rough experimental data; bars
represent the statistical errors. Lines: many body Coulomb trajectory
calculations for a volume of $\sim 3V_0$ (full), $4V_0$ (dotted), $5V_0$ 
(dashed dotted) and $6V_0$ (long dashed) within the two extreme freeze
out hypotheses.}
\label{fig:10}
\end{figure}
A parameter to be discussed is the freeze-out volume, which determines the
average value of the Coulomb potential energy. As already remarked
in section 6, the Coulomb energy is positively correlated to the fragment
multiplicity implying that the effect of the volume will depend on the
freeze-out hypothesis. If we change in an arbitrary way the freeze out
volume as well as the other characteristics of the freeze-out configuration,
the quantitative result for the heat capacity varies considerably, though
only very extreme and unrealistic hypotheses suppress the negative 
branch~\cite{bormio_michela}. However it is possible to make a few steps 
further.
In fact the average volume can be experimentally estimated in a quite
precise way from the mean detected fragment kinetic energy ($<E_{kin}>$) 
of the data sample under study. 
Concerning the QP data, the measured fragment energy distributions are 
compatible with a collective (i.e non thermal, non Coulomb) component at 
most 0.7 A.MeV at 7 A.Mev of excitation energy, and about 0.4 A.MeV at 
E*=4 A.MeV~\cite{michela}. 
This component has been subtracted from the presented data and all the
results presented in this paper would not change in any sizeable way if this
component was kept. Therefore for this data sample one can perform 
a many body Coulomb trajectory
calculation, by randomly placing the reconstructed primary fragments in a
spherical volume and letting them evolve in the Coulomb field. 
Under the hypothesis that, on average, light charged particle evaporation does 
not affect fragment velocities, the superposition of the average Coulomb and
thermal motion provides an observable directly comparable with $<E_{kin}>$.
This comparison should allow to select directly from data a (possibly energy
dependent) range of freeze-out volumes. The result is displayed in figure 10
for the quasi-projectile data in the two extreme freeze out hypotheses. The
full lines correspond to the smallest volumes that contain the fragments.
This minimum volume turns out to be on average $\sim 3V_{0}$. For each bin
energy the temperature, determining the average thermal motion $1.5\cdot T$, 
is univocally determined from energy conservation, using the kinetic
equation of state eq.(\ref{temp}). Of course the actual value of the
temperature depends also on the level density parameter entering equation 
(\ref{temp}). This extra source of uncertainty however does not modify in any
sizeable way the results. In fact the Coulomb contribution by far dominates
the average kinetic energies, implying that the average freeze out volume
can be estimated independently of the level density parameter.

Once the freeze out hypothesis is fixed, the results of figure 10 define
unambiguously the average volume. From this comparison the 
{\it cold fragment hypothesis} is ruled out, because the average
Coulomb energy per fragment turns out much smaller than in the data. We note
by passing that the good reproduction of the measured average kinetic
energies can be considered an additional evidence of the
equilibration of the data sample. Any collective component or dynamical
mechanism in fragment formation would lead to a deviation between the
detected energies and the ones reconstructed through Coulomb trajectories;
this discrepancy would increase with increasing deposited energy.

The situation is slightly more complicated in the case of central {\it Xe+Sn}
collisions. In this case $<E_{kin}>$ gives a measure of the Coulomb repulsion plus
the radial collective flow. In the hot fragment hypothesis a volume of 
$3V_{0}$ is consistent with experimental data if a collective flow of 0.6
A.MeV is assumed, but a volume as large as $6V_{0}$ can still reproduce the
fragment kinetic energy if the collective flow is 1 A.MeV\cite{remi} for
32 A.MeV bombarding energy. One
could also consider smaller freeze out volumes and lower
collective components, as suggested by some theoretical 
speculations~\cite{campi_cmd}. 
\begin{figure}[tbh]
\begin{center}
\epsfig{figure=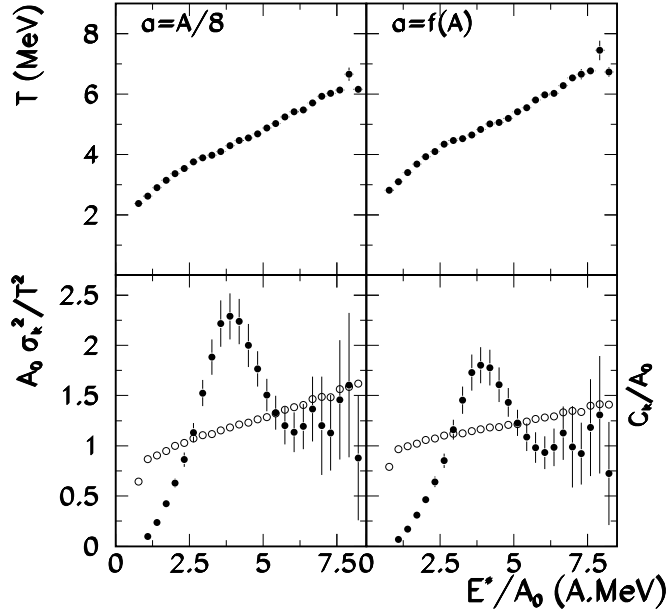,width=10.cm}
\end{center}
\caption{\it Temperature, kinetic energy fluctuation and kinetic heat capacity
for the QP data in the hot fragment hypothesis (freeze-out volume $\sim
3V_0$ with two different prescriptions for the level density parameter
(for $f(A)$ see Ref.~\protect\cite{lev_dens}).}
\label{fig:11}
\end{figure}

The last unknown entering the total heat capacity eq.(\ref
{eqcentral}) is the level density parameter $a$ that determines the freeze
out temperature. The effect of changing $a$ is illustrated in figure 11 for
the quasi-projectile data: the net effect is an uncertainty in the
estimate of the temperature that never exceeds about 0.5 MeV per nucleon.
If the abnormal fluctuation signal is clearly independent of the detailed
structure of the level density parameter, the localization of the
divergences however depends on the value assumed. It is important to keep in
mind that in any case the latent heat cannot be quantitatively estimated, 
because of the fluctuation suppression operated by our reconstruction method. 
This means that the distance between the two crossing points between 
$\sigma_k^2/T^2$ and $c_k$ in figure 11 represents in both cases a lower
limit for the actual value of the latent heat.

\begin{figure}[htb]
\begin{center}
\epsfig{figure=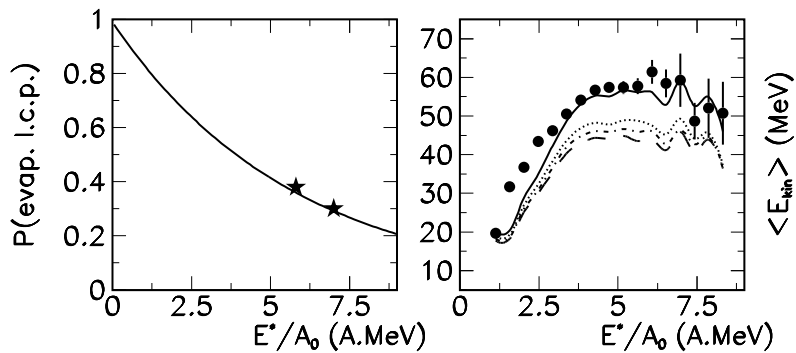,width=12.cm}
\end{center}
\caption{\it Left part: percentage of evaporated light charged particles as a
function of the excitation energy; symbols: experimental data from 
Ref.~\protect\cite{abdou}; line: exponential extrapolation. Right part: 
as figure 10, but lines refer to the freeze out reconstruction hypothesis 
defined in the left panel.}
\label{fig:12}
\end{figure}

The study of the measured $<E_{kin}>$ has allowed us to exclude the cold fragment
hypothesis for the reconstruction of the freeze out. However the hot
fragment hypothesis is also an extreme scenario for the secondary
deexcitation. It is reasonable to expect that the percentage of primary
light particles is neither 0\% nor 100\% and rather depends on the deposited
energy. A more quantitative insight into the thermodynamical properties of the
hot nuclear systems can be achieved if experimental information about the
freeze out composition is inserted in the freeze-out reconstruction. An
example is given by Ref.~\cite{abdou}. Velocity correlations between 
fragments and light charged particles allow to estimate the percentage of
secondarily evaporated particles (symbols in left panel of Fig.12) and the
average excitation energy of primary fragments (stars in the right panel of
Fig.13). This work~\cite{abdou} has been performed on the same 
{\it Xe+Sn} sample
considered in the present paper. We have already mentioned that the size of
the sources measured in central {\it Xe+Sn} and in the {\it Au+Au} peripheral 
events are very similar.  If we assume that for a given excitation energy the
breakup of the sources does not depend on the entrance channel, we can
perform for the quasi-projectile events a freeze-out reconstruction based on
the results of Ref.~\cite{abdou}.

Within this freeze out assumption it is possible to reproduce the QP
measured kinetic energies as in figure 10 with a freeze out volume 
$\sim 3V_0$ (corresponding to the minimum volume that contains the fragments).
This is shown in the right part of figure 12. The $<E_{kin}>$ observable is quite
well reproduced by the superposition of Coulomb and thermal motion. Only at
energies lower than about 3 A.MeV the kinetic energies are underestimated.
This is clearly due to the arbitrariness of our low energy extrapolation.
While it is reasonable to assume that the totality of light charged
particles is evaporated at an excitation energy close to zero, a comparison
between figure 12 and figure 10 suggests that the percentage of evaporated
light charged particles in the interval $0 < E^{*}/A_{0} < 3$ A.MeV should be
less steep. Some data in this energy domain are clearly needed~ to better
constrain the curve.

\begin{figure}[htb]
\begin{center}
\epsfig{figure=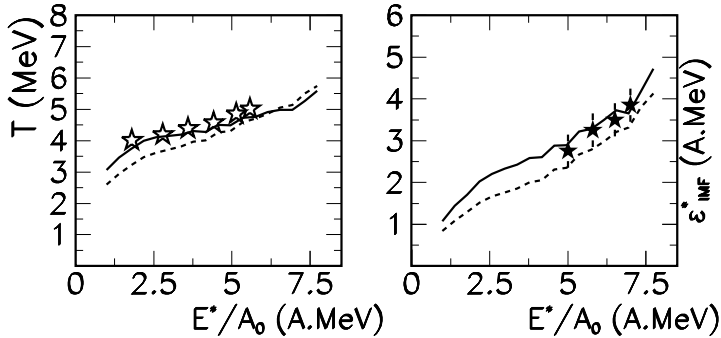,width=12.cm}
\end{center}
\caption{\it Left: temperature measured with an isotopic 
thermometer~\protect\cite{prctrieste} (open symbols)
and from eq.(\ref{temp}) for the QP (lines). 
Right: primary fragment internal excitation energy measured from
velocity correlation measurements in central collisions~\protect\cite{abdou} 
(full symbols) and from eq.(\ref{temp}) for the QP (lines). Dashed
lines correspond to a level density parameter $a=A/8$ while $a=f(A)$ (see
Ref.~\protect\cite{lev_dens}) is taken for the full lines.}
\label{fig:13}
\end{figure}

The last quantity to be settled in the heat capacity analysis is the level
density parameter. This can be fixed by injecting as much experimental
information as possible from independent measurements. As an example, symbols
in the left panel of figure 13 show the isotopic temperatures from the Carbon
thermometer measured for the quasi projectile data~\cite{prctrieste}. For
this specific thermometer side feeding effects have been estimated to induce
an uncertainty of about 0.5 MeV at most~\cite{prctrieste}. Symbols in the 
right panel of figure 13 show the internal excitation energy of primary 
fragments, experimentally reconstructed for the central {\it Xe+Sn} data in 
Ref.~\cite{abdou}. Both sets of data are compared in figure 13 with the kinetic
thermometer eq.(\ref{temp}) measured for the quasi projectile data sample. 
The results of figure 13 do not allow to discriminate between
the two different prescriptions for the level density parameter, however
they indicate that  the temperature cannot vary more than what shown by
figure 11.

\begin{figure}[tbh]
\begin{center}
\epsfig{figure=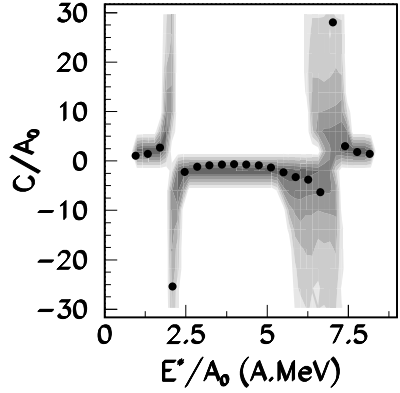,width=10.cm}
\end{center}
\caption{\it Heat capacity per nucleon as a function of the excitation 
energy for the QP system with the freeze out reconstruction constrained to 
reproduce the experimental values of  Ref.~\protect\cite{abdou}.
A level density $a=A/8$ has been assumed.}
\label{fig:14}
\end{figure}

\begin{figure}[tbh]
\begin{center}
\epsfig{figure=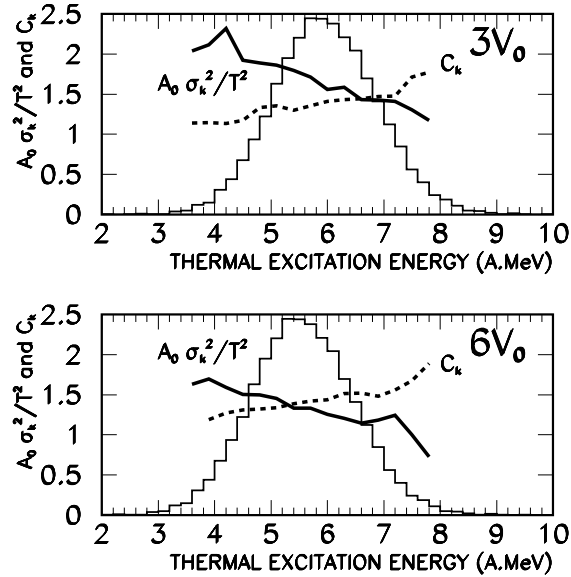,width=10.cm}
\end{center}
\caption{\it Normalized fluctuations (full lines), kinetic heat capacity (dashed
lines) and excitation energy distribution (histograms)  for the central 32 A.MeV
{\it Xe+Sn} system at the lowest and highest freeze out volume with the freeze out
reconstruction constrained to reproduce the experimental values of 
Ref. \protect\cite{abdou}. A level density $a=A/8$ has been assumed.}
\label{fig:15}
\end{figure}

The final result for the heat capacity measured in the two Au-like systems
is presented in figures 14 and 15. In both cases the freeze out volume and
multiplicity have been fixed through the independent experimental
constraints shown in figures 12 and 13. In the case of central collisions
(figure 15) only a limiting range of possible freeze out volumes 
($3V_{0} < V < 6V_{0}$) has been estimated from the analysis of asimptotic 
$<E_{kin}>$, because of the 
additional uncertainty coming from the radial collective 
flow~\cite{remi,nicolas}. The compatibility of the two sets of data and the 
presence of divergences and a negative branch for $c$ is indisputable.

\section{Conclusions}

The sources of uncertainty that arise in the thermostatistical analysis of
well detected multifragmenting nuclear systems have been analyzed. In order
to be able to study mean values as well as higher moments of the
distributions, the quality of the calorimetric reconstruction is essential.
A general protocol has been established to deal with missing information in
such a way that the distortions due to an imperfect detection are minimized.
The presence of a first order liquid-to-gas-like phase transition in nuclear
mul\-ti\-frag\-men\-ta\-tion~\cite{plb,remi} is confirmed by this analysis.
The negative heat capacity signal survives to all the uncertainties due to
the different reconstruction hypotheses. A negative value for the heat
capacity is signed by abnormally large interaction energy fluctuations. 
In order to disentangle between physical fluctuations and experimental
uncertainties we have systematically adopted a procedure which suppresses
the variance of all not measured quantities. The challenge for the next
future is to reintroduce this missing fluctuation by  more complete
measurements and minimum bias simulations. A promising technique to recover
the missing information is suggested by figure 10 above. More sophisticated
Coulomb trajectories can be employed to compare the variance of the
asymptotic fragment kinetic energy ($<E_{kin}>$) distribution. Fluctuations at
freeze out could then be tuned to reproduce this observable~\cite{inprogr}.

The ultimate challenge of these analysis is the reconstruction of the
nuclear phase diagram as a function of mass and possibly isospin. To this
aim as accurate as possible measurements of relevant thermodynamical
parameters at freeze out are essential. A lot of work in this direction has
been already done and will hopefully continue and become more precise with
next generation detectors~\cite{chimera}. With only two independent
measurements (for instance temperature and volume) at the 10\% or 20\% level,
almost all ambiguities in the quantitative estimate of the heat capacity
is removed. Any other independent observation of freeze out variables can
then be used as a cross check of the consistency of the procedure.

\ack
This work has been partially supported by NATO grants CLG 976861 and by
grants of the Italian Ministry of University and Scientific and
Technological Research under grants Cofin99. 


\begin{thebibliography}{99}
\bibitem{plb}  {\small M.~D'Agostino et al., Phys. Lett. B 473 (2000) 219.}

\bibitem{remi}  {\small N.~Le~Neindre et al., contribution to the XXXVIII
Winter Meeting on Nucl. Phys., Bormio (Italy), Ed. I. Iori and A. Moroni,
(2000) p.404 and to be published.}

\bibitem{gravity}  {\small W.~Thirring, Zeit. Phys. 235 (1970) 339;\newline
D.~Lynden-Bell, Proc. of the XXth IUPAP Int.Conf on Stat. Phys., Paris, July
20-24 (1998) and cond-mat/9812172.}

\bibitem{gross}  {\small D.~H.~E.~Gross, Phys. Rep. 279 (1997) 119;
D.H.E.~Gross and E. Votyakov, Europhys. Journ. B15 (2000) 115.}

\bibitem{prl99}  {\small F.~Gulminelli and P.~Chomaz, Phys. Rev. Lett. 82
(1999) 1402.}

\bibitem{schmidt}  {\small M.~Schmidt et al., Phys. Rev. Lett. 79 (1997) 99;
Nature 393, 238 (1998); Phys. Rev. Lett. 86 (2001) 1191.}

\bibitem{analytical}  {\small P.~Chomaz and F.~Gulminelli, Nucl.~Phys. A647
(1999) 153.}

\bibitem{michela}  {\small M.~D'Agostino et al., Nucl. Phys. A650 (1999)
329.}

\bibitem{xesn}  {\small N.~Marie et al., Phys. Lett. B391 (1997) 15.}

\bibitem{richertrep}  {\small J.~Richert and P.~Wagner, 
arXiv:nucl-th/0009023 v2 and to be published in Phys. Rep.}

\bibitem{botet}  {\small R.Botet, M.~Ploszajczak, Phys. Rev. E62 (2000) 1825
and nucl-ex/0101012.}

\bibitem{richert}  {\small J. M. Carmona, N. Michel, J. Richert and P.
Wagner, Phys. Rev. C61 (2000) 037304.}

\bibitem{lebowitz}  {\small J.~L.~Lebowitz et al., Phys.Rev. 153 (1967) 250.}

\bibitem{iso_iso}  {\small Ph.~Chomaz, F.~Gulminelli and V.~Duflot, Phys.
Rev. Lett. 85 (2000)3587.}

\bibitem{cugnon}  {\small J.~Cugnon and D.~L'Hote, Nucl. Phys. A397 (1983)
519.}

\bibitem{frankland}  {\small J.~D.~Frankland et al., contribution to the
XXXV Winter Meeting on Nucl. Phys., Bormio (Italy), Ed. I. Iori, (1997)
p.323.}

\bibitem{lecolley}  {\small J.~F.~Lecolley et al{\it .}, Nucl. Instr. and
Methods A441 (2000) 517.}

\bibitem{msuganil}  {\small C.~P.~Montoya et al{\it .}, Phys. Rev. Lett. 73
(1994) 3070;\newline
T.~Lefort et al{\it .}, Nucl. Phys. A662 (2000) 397;\newline
E.~Plagnol et al., Phys. Rev. C61 (2000) 014606.}

\bibitem{prctrieste}  {\small P.~M.~Milazzo et al{\it .}, Phys. Rev. C58
(1998) 953.}

\bibitem{bondorf}  {\small J.~P.~Bondorf, A.~S.~Botvina, A.~S.~Iljinov,
I.~N.~Mishustin, K.~Sneppen, Phys. Rep. 257 (1995) 133.}

\bibitem{elliot}  {\small J.~B.~Elliot and A.~S.~Hirsch, Phys. Rev. C61
(2000) 054605.}

\bibitem{charity}  {\small R.~J.~Charity et al., Nucl. Phys. A483 (1988)
371.}

\bibitem{alex}  {\small J.~P.~Bondorf, A.~S.~Botvina, I.~N.~Mishustin, Phys.
Rev. 58C (1998) R27.}

\bibitem{bormio_michela}  {\small M.~D'Agostino et al., contribution to the
XXXVIII Winter Meeting on Nucl. Phys., Bormio (Italy), Ed. I. Iori and A.
Moroni, (2000) p.386.}

\bibitem{pierre}  {\small P.~D\'{e}sesquelles, M.~D'Agostino, A.~S.~Botvina,
M.~Bruno {\it et al.}, Nucl. Phys. A633 (1998) 547.}

\bibitem{lev_dens}  {\small The level density parameter is assumed, as in
Ref.~\cite{bondorf}, continuously varying from $A/8$ for the lightest
nucleus $A=7$ to $A/12$ for the heaviest one $A=197$, to take into account
different surface and volume contributions.}

\bibitem{epax}  {\small K.~S\"{u}mmerer, W.~Bru\"{u}chle, D.~J.~Morrissey,
M.~Sch\"{a}del, B.~Szweryn and Yang Weifan, Phys. Rev. C 42 (1990) 2546.}

\bibitem{eos}  {\small J.~A.~Hauger, P.~Warren, S.~Albergo, F.~Bieser and
EOS collaboration, Phys. Rev. C {\bf 57} (1998) 764 and references quoted
therein.}

\bibitem{campi_cmd}  {\small X. Campi et al., Nucl. Phys. A681 (2001) 458c;
\newline
A.~Chernomoretz, M.~Ison, S.~Ortiz, C.~Dorso, nucl-th/0101061, 26 Jan. 2001.}

\bibitem{abdou}  {\small N.~Marie, A.~Chbihi, J.~B.~Natowitz, A.~Le
F\`{e}vre {\it et al.}, Phys. Rev. C58 (1998) 256; \newline
S.~Hudan et al.,
contribution to the XXXVIII Winter Meeting on Nucl. Phys., Bormio (Italy),
Ed. I. Iori, (2000) p.404.}

\bibitem{nicolas}  {\small N.~Le~Neindre, These de l'Universit\'{e}
de Caen (1999) unpublished.}

\bibitem{inprogr}  {\small M.~D'Agostino et al., in preparation.}

\bibitem{chimera}  {\small A.~Pagano et al. Nucl. Phys. A681 (2001) 331c.}
\end{thebibliography}
\end{document}